\documentclass[%
 reprint,
showpacs,
showkeys,
 amsmath,amssymb,
 aps,
]{revtex4-1}

\usepackage{graphicx}
\usepackage{dcolumn}
\usepackage{bm}
\usepackage{hyperref}

\newcommand{\be}{\begin{displaymath}}
\newcommand{\ee}{\end{displaymath}}
\newcommand{\ba}{\begin{eqnarray*}}
\newcommand{\ea}{\end{eqnarray*}}

\begin{document}


\title{Inference of plasmid-copy-number mean and noise\\
from single-cell gene expression data}

\author{St\'ephane Ghozzi}
 \altaffiliation[now at ]{Institut f\"ur Theoretische Physik, Universit\"at zu K\"oln, Z\"ulpicherstrasse 77, 50937 Cologne, Germany}
  \email{ghozzi@thp.uni-koeln.de}

\author{J\'er\^ome \surname{Wong Ng}}
 \altaffiliation[now at ]{Physics of Biological Systems, CNRS URA 2171, Institut Pasteur, 25-28, rue du Dr Roux, 75015 Paris, France.}
\affiliation{Laboratoire de Physique Statistique, \'Ecole Normale Sup\'erieure, UPMC Univ Paris 06, Universit\'e Paris Diderot, CNRS, 24 rue Lhomond, 75005 Paris,  France.}

\author{Didier Chatenay}
\author{J\'er\^ome Robert}
 \affiliation{Laboratoire Jean Perrin, FRE 3231 CNRS-UPMC, 24 rue Lhomond, 75005 Paris, France.
}

\date{\today}

\begin{abstract}
Plasmids are extra-chromosomal DNA molecules which code for their own replication. We previously reported a setup using genes coding for fluorescent proteins of two colors that allowed us, using a simple model, to extract the plasmid copy number noise in a monoclonal population of bacteria [J. Wong Ng et al., Phys. Rev. E, 81, 011909 (2010)]. Here we present a detailed calculation relating this noise to the measured levels of fluorescence, taking into account all sources of fluorescence fluctuations: the fluctuation of gene expression as in the simple model, but also the growth and division of bacteria, the non-uniform distribution of their ages, the random partition of proteins at divisions and the replication and partition of plasmids and chromosome. We show how using the chromosome as a reference helps extracting the plasmid copy number noise in a self-consistent manner.
\end{abstract}


\maketitle

\section{Introduction}

Plasmids are highly common in natural bacterial strains and are widely used in studies
of gene expression~\cite{Solar1998}. They have been seen as a model for genomic replication and partition~\cite{Solar1998,Nordstroem2006} and studied as genetic control systems, possibly subject to noise~\cite{Paulsson2001}. A number of technics have been used to measure plasmid copy numbers (PCN). DNA titration is the simplest, but least precise. Quantitative polymerase chain reaction (qPCR)~\cite{Lee2006} is often used and gives access to mean PCN in a population. Two \emph{in vivo} labeling techniques may \emph{a priori} give access to PCN distributions when applied on single-cells: fusions of a fluorescent protein with a transcription factor that binds the plasmids~\cite{Belmont2001,Pogliano2001} or insertion of a gene coding for a fluorescent protein into the plasmids~\cite{Bagh2008}. However both have limitations that prevent them from giving access to more than the mean PCN~\cite{WongNg2010}.

In the remainder of this Introduction we briefly recall the setup of the experiments reported previously, making use of dual fluorescence reporters, that allowed us to infer the second moment of PCN distributions~\cite{WongNg2010}. In Section \ref{SimpleModel} we derive the expression for PCN mean and noise in a simple case, where only fluctuations of gene expression are considered. The realistic case, taking into account all sources of fluctuations of the actual experiment, is presented in Section \ref{CompleteModel}. Section \ref{Results} presents the values obtained for PCN mean and noise when one uses the experimentally measured quantities. These results and the principle of this work are then discussed. Appendixes present some computations in greater details.

The gene {\it egfp} ~\cite{Tsien1998}, coding for the the green fluorescent protein EGFP, was fused to the inducible, strong promoter {\it PtacI} {\cite{Boer1983} and then inserted in the chromosome of an {\it E. coli} strain. The bacteria were then transformed with either one of the four plasmids studied here, which contained the fusion {\it PtacI-mOrange} {\cite{Shaner2004}: we thus obtained strains expressing EGFP and the orange fluorescent protein mOrange at the same time, under the same transcriptional control. After one hour  induction with IPTG, all protein expression was blocked. Cells were incubated overnight so that all fluorescent proteins acquire their mature form.  For each of the four strains, green and orange fluorescence intensities of individual cells were then measured. In each experiment at least 10,000 cells were observed, and at least three experiments were done in each condition.

In general, disentangling the various contributions to the final distribution of fluorescence would be a difficult problem. However, making some assumptions on the gene expression processes, we will be able to express the first and second moments of the number of fluorescent proteins as functions of those of copy numbers and to inverse these relations to find how to relate the experimental measurements to the distribution of PCN. The next section presents this strategy in a simple case. 
\begin{figure*}
\begin{center}
\includegraphics[scale=0.35]{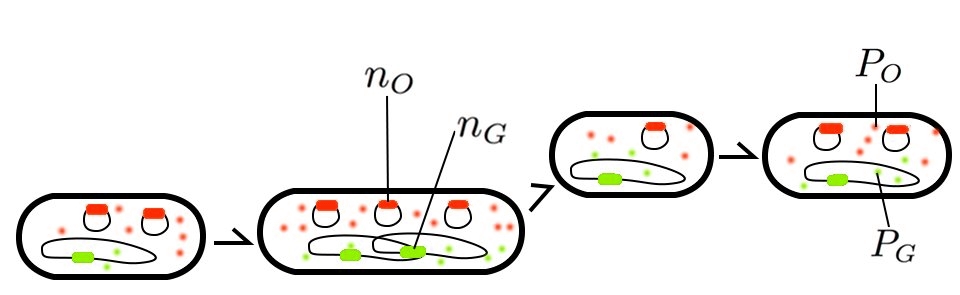}
\caption{(Color online) Cartoon of the lineage of a bacterium during protein production induction, here depicted with one division (only one of the two final cells is shown). Fluorescence intensities of single cells are measured at the end of induction. The orange, resp. green, intensities are proportional to the number of orange proteins $P_O$, resp. green proteins $P_G$, in the observed cell, shown as orange (dark gray), resp. green (light gray), dots. These proteins were produced during all the induction by a varying number of {\it mOrange} or {\it egfp} copies ($n_O$ and $n_G$) and randomly distributed among daughter cells at each division.}
\label{FigExp}
\end{center}
\end{figure*}

\section{Simple model}
\label{SimpleModel}

We suppose here that during the induction, bacteria do not grow, the plasmids and chromosomes do not replicate, the protein production does not depend on time \footnote{This hypothesis is not necessary, supposing that it does not depend on time \emph{on average} would lead to the same result, but it makes the notations simpler.} and the age distribution of bacteria is uniform.

We note $P_a^i$ the contribution of the copy $i$ of the gene $a$ ($a=O$ or $G$
for the genes {\it mOrange} or {\it egfp}) to the total number of proteins $P_a$ at the end of induction in one cell and 
$n_a$ the number of copies of the gene $a$ in that cell (see Fig.~\ref{FigExp}). One can write:
\be
P_a =\sum_{i=1}^{n_a}P_a^i.
\ee
The average (over the population) of $P_a$ can thus be written:
\be
\langle P_a\rangle = \sum_{n_a}\sum_{i=1}^{n_a}\sum_{P_a^i}p(n_a, P_a^i)\, P_a^i,
\ee
where $p(n_a, P_a^i)$ is the joint probability of $n_a$ and $P_a^i$. We can suppose that the distribution of the number of proteins produced by each copy does not depend on the particular copy considered nor on the number of copies (we measured the same distributions of green fluorescence, i.e. of expression from the chromosome, for strains bearing both high and low copy number plasmids~\cite{WongNg2008}). Thus:
\ba
\langle P_a\rangle & = & \sum_{n_a}p(n_a)\, n_a\sum_{P_a^1}p(P_a^1)\, P_a^1 \\
& = & \langle n_a\rangle\langle P_a^1\rangle.
\ea
Moreover we can suppose that on average the number of proteins produced by a copy of a gene does not depend on the gene (both genes are under the same promoter). Hence, as expected:
\begin{equation}
\frac{\langle n_O\rangle}{\langle n_G\rangle} = 
\frac{\langle P_O\rangle}{\langle P_G\rangle}.
\label{EqAvSimple}
\end{equation}
The moments of order 2 can similarly be written:
\be
\langle P_aP_b\rangle = \sum_{n_a, n_b}
\sum_{i=1}^{n_a}\sum_{j=1}^{n_b}\sum_{P_a^i, P_b^j}p(n_a, n_b, P_a^i, P_b^j)
\, P_a^i P_b^j.
\ee
where $P_a$ and $P_b$ are evaluated in the same cell.

In the case of different genes, we can suppose that the correlation does not depend on the particular copies considered, nor on their numbers. Thus: 
\ba
\langle P_OP_G\rangle & = &\!\! \sum_{n_O, n_G}\!\! p(n_O, n_G)\, n_O n_G 
\!\!\!\sum_{P_O^1, P_G^1}\!\! p(P_O^1, P_G^1)\, P_O^1 P_G^1\\
& = & \langle n_On_G\rangle\langle P_O^1P_G^1\rangle.
\ea
In the case of the same gene, we can suppose that two different copies correlate like two copies of different genes ($\langle P_a^iP_a^j\rangle = \langle P_O^1P_G^1\rangle,\;\forall i\neq j$) and that the auto-correlation of one copy does not depend on the particular copy or gene considered ($\langle (P_a^i)^2\rangle = \langle (P^1)^2\rangle,\;\forall a, i$). Then:
\be
\langle P_a^2\rangle = \langle n_a\rangle\langle (P^1)^2\rangle
+ \langle n_a(n_a -1)\rangle\langle P_O^1P_G^1\rangle.
\ee
Combining those two last expressions with equation \ref{EqAvSimple}, we obtain:
\be
\langle n_O^2\rangle\! =\! \frac{\langle P_O\rangle}{\langle P_G\rangle} 
\langle n_G^2\rangle + \frac{1}{\langle P_OP_G\rangle}\!\! \left(\!\! \langle P_O^2\rangle
\!-\!\frac{\langle P_O\rangle}{\langle P_G\rangle}\langle P_G^2\rangle\!\! \right)
\!\! \langle n_On_G\rangle.
\ee
Since the replication of the chromosome is well controlled~\cite{Skarstad1986,Nordstroem2006} 
we can suppose
that the variance of the chromosome copy number vanishes ($\langle n_G^2\rangle\approx\langle n_G\rangle^2$) and that the plasmid and chromosome copy numbers are uncorrelated
($\langle n_On_G\rangle\approx\langle n_O\rangle\langle n_G\rangle$). Let $\eta$ be the PCN noise, defined by: $\eta^2=(\langle n_O^2\rangle-\langle n_O\rangle^2)/\langle n_O\rangle^2$. Then:
\begin{equation}
\eta^2=\frac{\langle P_G\rangle}{\langle P_O\rangle} 
+ \frac{1}{\langle P_OP_G\rangle}
\!\left(\!\frac{\langle P_G\rangle}{\langle P_O\rangle}\langle P_O^2\rangle
-\langle P_G^2\rangle\!\right)\!-\!1,
\end{equation}
which, it turns out, does not depend on the chromosome copy number or any other external inputs, but solely on quantities directly measured in this experiment.

\section{Complete model}
\label{CompleteModel}

We want now to also take into account sources of fluorescence fluctuation other than gene expression. We assume that \emph{all cells have exactly the same division time $T$}. Two studies report a small variability of division times, with a standard deviation of the growth time constant of $\sim$10\% of the average~\cite{Megerle2008,Wang2010}. 
We note $t_0$ the age of a cell at the beginning of induction. Under this hypothesis, the distribution of ages $t_0$ is exponential~\cite{Neidhart1996}: $p(t_0)=(2\ln2/T).2^{-t_0/T}$. We will also consider that the induction time (one hour) is a multiple of the division time. This is true at $30$ and $37\,^{\circ}\mathrm{C}$, where we measured cell cycles of 1 h and 30 min respectively, but not for intermediate temperatures (this is discussed in Section \ref{Results}). We will present calculations with cells dividing twice during the induction, i.e. a cell cycle of 30 min; more or less divisions only change the numerical pre-factors~\cite{Ghozzi2009}.  

At each cell division, fluorescent proteins are randomly inherited by one of the two daughter cells, thus adding to the fluorescence fluctuations. 
As discussed in Appendix~\ref{App:Div}, this contribution turns out to be small: to a good precision, half of the fluorescent proteins are inherited by each daughter cell. 

Following one lineage during the induction, we can now express the number of fluorescent proteins at the end of induction in a given cell:
\be
P_a=\left(\frac{1}{4}\int_{t_0}^T +\frac{1}{2}\int_{T}^{2T} +\int_{2T}^{2T+t_0}\right)\!\!\sum_{i=1}^{n_a(t)}  \alpha_a(i,t)\,dt,
\ee
where we took the age of the cell at the beginning of induction  $t_0$ as the initial time and introduced $\alpha_a(i,t)$, the rate of protein production at time $t$ from the copy $i$ of the gene $a$ \footnote{Here both $\alpha_a$ and $n_a$ are particular realizations of rate and copy number, and thus are ``noisy'' functions. One can fix an arbitrarily small time step and consider the \emph{initiations of transcription} to give a precise, well defined meaning to $\alpha_a$ without disregarding the discrete replication, transcription, translation and maturation steps.}. 

\subsection{Fluorescence averages}

To compute the average of $P_a$ we introduce the joint probability $p[t_0,n_a,\alpha_a]$, which is now a functional and the integral is performed over all possible $n_a$ and $\alpha_a$ functions: 
\begin{equation}
\langle  P_a \rangle = \int dt_0 \mathcal{D}[n_a] \mathcal{D}[\alpha_a]\; 
p[t_0,n_a,\alpha_a]\, P_a[t_0,n_a,\alpha_a]. \label{Eq:Av1}
\end{equation}
A number of assumptions on gene expression and replication, similar to those presented in the Section~\ref{SimpleModel}, are detailed in Appendix~\ref{App:Assump}. We use here the hypotheses \emph{(i)} to \emph{(iv)} to simplify Eq.~\ref{Eq:Av1} \emph{without having to postulate explicit models for gene expression or replication}. We then find:
\be
\langle P_a\rangle=\frac{3}{4}\, T \langle  \alpha \rangle   \left( \langle\overline{n_a}\rangle + \frac{1}{T}\int_0^T\!\!dt_0\, p(t_0) \int_0^{t_0}\!\!dt\, \langle  n_a(t) \rangle \right), 
\ee
 where $\overline{\bullet}$ is the average over one cycle, which commutes with the average over the population.

In general we cannot inverse this relation so as to express the average copy number as a function of the average protein number and we do not know the plasmid replication systems well enough to evaluate the second term in the parentheses. It is nevertheless possible to bound its ratio to the mean copy number. We thus define $\mathcal{R}_a = ((1/T)\int_0^T dt_0\, p(t_0) \int_0^{t_0} dt\,\langle  n_a(t) \rangle)/\langle\overline{n_a}\rangle$, and use it to express the mean PCN per chromosome:
\begin{equation} \label{nOnG}
\frac{\langle\overline{n_O}\rangle}{\langle\overline{n_G}\rangle} =\left(\frac{1+\mathcal{R}_G}{1+\mathcal{R}_O}\right)\frac{\langle P_O\rangle }{\langle P_G\rangle }.
\end{equation}
We show in Appendix~\ref{App:RST} that $\mathcal{R}_a \in [0.15,0.45]$. We also computed it after postulating various shapes for $\langle  n_a \rangle$ as a function of time and propose that this interval can be reduced to $[0.36,0.44]$ (see Appendix~\ref{App:Test}). The results for the four plasmids we studied, at various temperatures, are presented in Section \ref{Results}.

\subsection{Fluorescence cross-correlations}

We will follow the same strategy for the correlations, namely bound terms related to plasmid or chromosome replication and partition. Beside those already mentioned, we use the assumptions \emph{(v)} and \emph{(vi)} presented in Appendix~\ref{App:Assump} and introduce: 
\be
\mathcal{S}_{ab}=\frac{1}{\langle\overline{n_a}\;\overline{n_b}\rangle}\frac{1}{T^2}\!\!\int_0^T\!\!\!dt_0\, p(t_0)\!\!\int_0^{t_0}\!\!\!dt\!\!\int_0^{t_0}\!\!\!dt' \langle n_a(t)n_b(t')\rangle.
\ee
We can now write: 
\begin{equation} \label{POPG}
\langle P_OP_G\rangle  = \frac{9}{16}T^2\langle \alpha_O\alpha_G\rangle 
(1+\mathcal{R}_O +\mathcal{R}_G + \mathcal{S}_{OG}) 
\langle\overline{n_O}\rangle\langle\overline{n_G}\rangle,
\end{equation}
where $P_O$ and $P_G$ are evaluated in the same cell.
We show in Appendix~\ref{App:RST} that $\mathcal{S}_{OG}\in [0,0.45]$, and argue that this interval can be reduced to $[0.20,0.28]$ (see Appendix~\ref{App:Test}).

\subsection{Fluorescence auto-correlations}

We consider now the moment of order 2 for the same gene, i.e. $\langle P_a^2\rangle$, with $a=O$ or $G$. 
We make two more assumptions, \emph{(vii)} and \emph{(viii)} in Appendix~\ref{App:Assump}, and note $C_\alpha(|t-t'|)$ the auto-correlation function at two times $t$ and $t'$ of the rate of fluorescent protein production $\alpha$.

Our guess is that the results will not be affected by the particular form this auto-correlation function will take; to test it we will make two extreme hypotheses: (A) of very short ``memory'', (B) of infinite (over the whole induction time) ``memory''.

In the hypothesis (A), we suppose that after a very short time $\tau$ the expression of a copy of a gene correlates with itself the same way it correlates with other copies. This makes sense if $\tau$ is small compared to the replication time; and indeed, we expect a particular copy auto-correlation to stem from multiple translations of a given mRNA, which has a typical life time of the order of the minute in bacteria, or from transcriptional bursts, which were shown to happen over short time scales~\cite{Golding2005}. In contrast, genes are on average replicated once per cell cycle, i.e. every few tens of minutes.

We consider in this hypothesis that $C_\alpha$ is a peaked function at 0, with a non-zero value beyond a small time $\tau$ such that it does not depend on wether a previous copy was the ancestor of the considered copy or not :
\ba
C_\alpha^{\rm A}(|t-t'|) & = & \langle  \alpha^2 \rangle\times\tau\delta(t-t')\\
&&+\langle  \alpha_O\alpha_G\rangle(1-\tau\delta(t-t')) -\langle\alpha\rangle^2, 
\ea
This gives:
\begin{eqnarray}
\langle P_a^2\rangle_{\mathrm{A}}
&=& \frac{9}{16}T^2\langle \alpha_O\alpha_G\rangle  (1+2\mathcal{R}_a + \mathcal{S}_{aa}) \langle(\overline{n_a})^2\rangle\nonumber\\
&+& \frac{5}{16}\tau T(\langle \alpha^2\rangle\!-\!\langle \alpha_O\alpha_G\rangle )
(1\!+\!3\mathcal{R}_a)\langle\overline{n_a}\rangle. \label{P2A}
\end{eqnarray}

In the hypothesis (B), we suppose that $C_\alpha$ is  constant: 
\be
C_\alpha^{\rm B}(|t-t'|)=\langle  \alpha^2 \rangle-\langle\alpha\rangle^2.
\ee 
(We expect the actual form of $C_\alpha$ to be intermediate between those two, namely a smooth declining function on a time scale of a few minutes.)
The hypothesis (B) is less realistic. It could correspond to mutations distinguishing different copies of a given gene. By noting that at any previous time each copy has exactly one ancestor, this translates in:
\ba 
\sum_{i=1}^{n_a(t)}\sum_{i'=1}^{n_a(t')}\!\!\langle\alpha_a(i,t)\alpha_a(i',t')\rangle_{\mathrm{B}}=\langle \alpha_O\alpha_G\rangle n_a(t)n_a(t')\\
+ (\langle \alpha^2\rangle\!-\!\langle\alpha_O\alpha_G\rangle )
(n_a(t)\theta(t\!-\!t')+n_a(t')\theta(t'\!-\!t)),
\ea
where $\theta$ is the Heaviside function.

We then introduce a third quantity, $\mathcal{T}_a$, which is defined in Appendix~\ref{App:RST}, and can be shown to lay in the interval $[0,9.9]$. (We will argue in Appendix~\ref{App:Test} that this interval can be reduced to $[5.7,6.1]$.) Then:
\begin{eqnarray}
\langle P_a^2\rangle_{\mathrm{B}} & = & \frac{9}{16}T^2\langle \alpha_O\alpha_G\rangle  
(1+2\mathcal{R}_a + \mathcal{S}_{aa}) \langle(\overline{n_a})^2\rangle\nonumber\\
& + & \frac{1}{8}T^2(\langle \alpha^2\rangle\!-\!\langle \alpha_O\alpha_G\rangle )(1\!+\!\mathcal{T}_a)\langle\overline{n_a}\rangle. \label{P2B}
\end{eqnarray}

The two hypotheses (A) and (B) thus only lead to different factors for the contribution of the average copy number \footnote{The ratio of these factors is $\frac{5}{2}\!\!\left(\!\frac{1+3\mathcal{R}_a}{1+\mathcal{T}_a}\!\right)\!\!\frac{\tau}{T}\in[0.01,0.2]$ if we consider that $\mathcal{R}_a$ and $\mathcal{T}_a$ are independent. (This interval is reduced to $[0.02,0.03]$ if we let these two quantities vary in the intervals found with a set of test functions, see Appendix~\ref{App:Test}.)}. This term is expected to be small, even in the hypothesis (B), where $1\!+\!\mathcal{T}_a$ can be of the order of 10: the numerical pre-factor cancels it, one can expect $\langle \alpha^2\rangle$ and $\langle \alpha_O\alpha_G\rangle$ to be of the same order of magnitude and, already for the plasmid of lowest copy number and for the chromosome, $\langle\overline{n_a}\rangle$ is significantly smaller than $\langle(\overline{n_a})^2\rangle$. Moreover, if we let $\tau$ tend to the time of induction $2T$, we recover terms of the same order of magnitude, thus suggesting a low sensitivity to the actual mathematical translation of the hypotheses.
The results will be presented and discussed with only the hypothesis (A), the more realistic, being considered; full computations with test functions confirmed that very close values for the PCN noise were found in the hypothesis (B) (data not shown).

\section{Results and discussion}
\label{Results}

By combining Eq.~\ref{nOnG}, \ref{POPG} and \ref{P2A} so as to eliminate the gene expression rates, and assuming that the replication of the chromosome is well controlled, we can now express the PCN noise: 
\begin{eqnarray}\label{Eq:eta}
\eta^2\!\! & = &\!\! \left(\frac{1+\mathcal{R}_O}{1+\mathcal{R}_G}\right)\!\!\left(\frac{1+3\mathcal{R}_O}{1+3\mathcal{R}_G}\right)\!\!\left(\frac{1+2\mathcal{R}_G+\mathcal{S}_{GG}}{1+2\mathcal{R}_O+\mathcal{S}_{OO}}\right)\!\!\frac{\langle P_G\rangle}{\langle P_O\rangle}\nonumber\\
&&\!\! \!\! \!\! \!\! \!\! \!\! -1 +\! \left(\frac{1+\mathcal{R}_O+\mathcal{R}_G+\mathcal{S}_{OG}}{1+2\mathcal{R}_O+\mathcal{S}_{OO}}\right)\!\! \frac{1}{\langle P_OP_G\rangle}\times\nonumber\\
&&\!\! \!\! \!\! \!\! \!\! \!\!  \! \left\{\left(\frac{1+\mathcal{R}_O}{1+\mathcal{R}_G}\right)\!\! \frac{\langle P_G\rangle}{\langle P_O\rangle} \langle P_O^2 \right.\rangle\!\left. -\! \left(\frac{1+3\mathcal{R}_O}{1+3\mathcal{R}_G}\right)
\langle P_G^2\rangle\right\}.
\end{eqnarray}
Note that both the auto-correlation time $\tau$ introduced previously and the cell cycle length $T$ have also been eliminated. Only terms related to replication and partition of genes, which we can bound, and the experimentally measured moments of protein numbers remain \footnote{The number of proteins can be determined up to a constant multiplicative factor, the same for both colors (namely, the fluorescent intensity per EGFP molecule in the selected green channel)~\cite{WongNg2010}; here, the contribution of random partition at divisions having been neglected, only ratios of the same order show up, thus canceling this unknown factor.}. 
By making the conservative assumption that $\mathcal{R}_a$ and $\mathcal{S}_{ab}$ can independently take any value in their intervals, we can compute intervals in which the mean PCN per chromosome and the PCN noise are surely. They are presented in Table~\ref{tab:Res} 
\begin{table*}
\caption{\label{tab:Res}
Mean PCN per chromosome $\langle\overline{n_O}\rangle/\langle\overline{n_G}\rangle$ and PCN noise $\eta$ computed with data from experiments at $37\,^{\circ}\mathrm{C}$, using the simple model or the complete one, either with a set of test functions or within a general analysis. Only the hypothesis (A) of short ``memory'' was considered. We assumed that cells divided twice during the induction.}
\begin{ruledtabular}
\begin{tabular}{lcccc}
& mini-F & mini-R1-$par^-$ & mini-R1-$par^+$ & mini-ColE1\\
\hline
$\langle n_O\rangle/\langle n_G\rangle$ simple & 0.9 &7.2 & 6.5 & 87\\
$\langle\overline{n_O}\rangle/\langle\overline{n_G}\rangle$ complete/test & [0.84, 0.95] & [6.8, 7.7] & [6.1, 6.9] & [82, 93]\\
$\langle\overline{n_O}\rangle/\langle\overline{n_G}\rangle$ complete/general & [0.71, 1.13] & [5.7, 9.1] & [5.1, 8.2] & [69, 110]\\
\hline
$\eta \times 10^2$ simple &58 & 36 & 30 & 28\\
$\eta \times 10^2$ complete/test & [50, 67] & [25, 45] & [16, 39] & [13, 38] \\
$\eta \times 10^2$ complete/general & [0, 100] & [0, 74] & [0, 71] & [0, 68] \\
\end{tabular}
\end{ruledtabular}
\end{table*}
for experiments at $37\,^{\circ}\mathrm{C}$, and in the Fig.~\ref{fig:Av}
\begin{figure}
\begin{center}
\includegraphics[clip, viewport=160 275 440 510,scale=0.75]{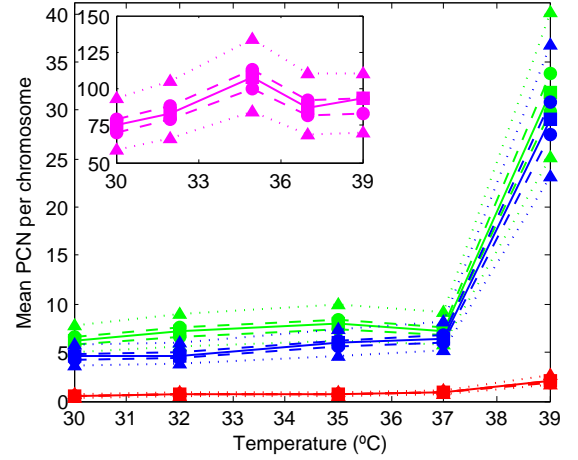}
\caption{\label{fig:Av} 
(Color online) Mean PCN per chromosome $\langle\overline{n_O}\rangle/\langle\overline{n_G}\rangle$, computed using 
Eq.~\ref{nOnG} and the measured average protein numbers $\langle P_O\rangle$ and $\langle P_G\rangle$. Results are shown for cells grown at $30$, $32$, $35$, $37$ and $39\,^{\circ}\mathrm{C}$ and for the four plasmids studied here, from bottom to top: mini-F (red), mini-R1-$par^+$(blue), mini-R1-$par^-$(green), mini-ColE1 (magenta). The values obtained in three cases are plotted: with the simple model (squares, solid line), with the complete model and test functions (upper and lower bounds of the interval: circles, dashed lines) or within a general analysis (upper and lower bounds of the interval: triangles, dotted lines). The mini-R1 plasmids used here have a synthetic, thermo-sensitive origin of replication, the control of which is inactivated at high temperature~\cite{Gerdes1985,Rodionov2004}.} 
\end{center}
\end{figure}
and Fig.~\ref{fig:Noise}
\begin{figure}
\begin{center}
\includegraphics[clip, viewport=135 250 460 535,scale=0.7]{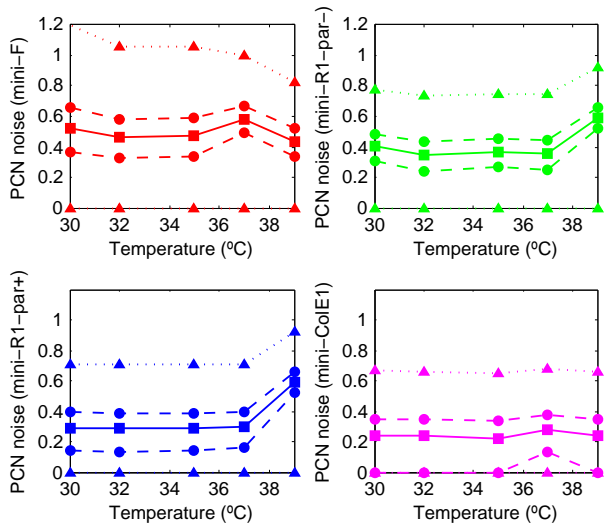}
\caption{\label{fig:Noise} 
(Color online) PCN noise $\eta$ computed using 
Eq.~\ref{Eq:eta}, and the measured average protein numbers $\langle P_O\rangle$, $\langle P_G\rangle$, and protein number correlations $\langle P_O^2\rangle$, $\langle P_G^2\rangle$, $\langle P_O P_G\rangle$. Results are shown at various temperatures for the four plasmids studied here (see the caption of Fig.~\ref{fig:Av}). We considered that cells divided once during the induction at $30$ and $32\,^{\circ}\mathrm{C}$, twice at $35$, $37$ and $39\,^{\circ}\mathrm{C}$. Only the hypothesis (A) of short ``memory'' was considered. The results obtained with the simple model are fully recovered if we suppose a similar behavior for the plasmids and for the chromosome, see the main text.}
\end{center}
\end{figure}
for various temperatures. We report both the intervals estimated with a general analysis and with a set of test functions for the moments of copy numbers. Values computed with the simple model are also shown. Both for means and noises, the values computed with the simple model fall in the middle of the intervals computed with the more realistic model. 

As Fig.~\ref{fig:Av} shows, we can clearly distinguish the plasmids by their mean PCN per chromosome. Moreover, these results agree with previous, independent estimates, as discussed~\cite{WongNg2010}. For the noises the picture is less clear. In the general study, the intervals found are too large for the results to be meaningful; but we know that we have largely overestimated them. In contrast, using test functions allows one to distinguish the plasmids by their PCN noises. In particular, we can notice that the partition system reduces the noise (compare mini-R1-$par^+$ and mini-R1-$par^-$), and that a plasmid with a high copy number (mini-ColE1) has a lower noise than a plasmid with a small copy number (mini-F), even though it has a partitioning system \footnote{We notice however that for the mini-R1 plasmids, both averages and noises increase at high temperature; this could come from fluctuations in the number of mature thermo-sensitive replication control proteins.}.

We tested the quality of the inference with simple computer simulations, where stochastic gene expression and plasmid replication were implemented (see Appendix~\ref{App:Sim} for more details). Table~\ref{tab:Sim} 
\begin{table*}
\caption{\label{tab:Sim}
Test of the inference method with computer simulations, in four cases: 1. a synchronized population of bacteria with fixed division time, equal to half the induction time; 2. as 1. with an exponential age distribution; 3. as 2. with a distribution of division times, with mean equal to half the induction time; 4. as 3. with the mean division time equal to one third of the induction time.}
\begin{ruledtabular}
\begin{tabular}{lcccc}
& case 1 & case 2 & case 3 & case 4\\
\hline
$\langle \overline{n_O}\rangle/\langle\overline{n_G}\rangle$ true & 11.9 & 9.6 & 10.1 & 10.1\\
$\langle n_O\rangle/\langle n_G\rangle$ simple & 12.8 & 10.1 & 11.0 & 11.7\\
$\langle\overline{n_O}\rangle/\langle\overline{n_G}\rangle$ complete/test & [12.0, 13.6] & [9.5, 10.7] & [10.4, 11.6] & [11.0, 12.4]\\
$\langle\overline{n_O}\rangle/\langle\overline{n_G}\rangle$ complete/general & [10.1, 16.1] & [8.0, 12.7] & [8.7, 13.8] & [9.3, 14.7]\\
\hline
$\eta \times 10^2$ true &63 & 66 & 75 & 75\\
$\eta \times 10^2$ simple &60 & 66 & 74 & 83\\
$\eta \times 10^2$ complete/test & [54, 67] & [59, 72] & [68, 80] & [77, 89] \\
$\eta \times 10^2$ complete/general & [0, 93] & [19, 98] & [35, 106] & [47, 114] \\
\end{tabular}
\end{ruledtabular}
\end{table*}
compares the true and inferred values of the mean PCN per chromosome and the PCN noise in four cases, corresponding to different assumptions on the age and cell cycle duration distributions. In each case we find a very good agreement.

As it appears in Eq.~\ref{nOnG}, \ref{POPG} and \ref{P2A}, what we call here ``plasmid copy number'', or ``chromosome copy number'', is precisely the average over one cell cycle of the number of copies of the gene coding for a fluorescent protein. A quantitative PCR (qPCR) measures $\langle n_O\rangle_{\mathrm{q}}/\langle n_G\rangle_{\mathrm{q}}$, where $\langle n_a\rangle_{\mathrm{q}}=\left\langle\int\!\!dt_0\,p(t_0)n_a(t_0)\right\rangle=\int dt_0\,p(t_0)\,\langle n_a(t_0)\rangle$. This quantity and the ratio $\langle\overline{n_O}\rangle/\langle\overline{n_G}\rangle$ reported here take in general different values. We have indeed noticed a discrepancy between the two approaches, but other explanations are likely~\cite{WongNg2010}.

We have made strong, but reasonable hypotheses on gene expression. We made intuitive notions explicit and gave them a well defined mathematical translation.

A deeper mathematical analysis could reduce significantly the general intervals found, but not below the intervals found with test functions.
Here again, the experimental approach and derivation of the PCN mean and noise are self-consistent: there are no external inputs, even in the bounded ``correction'' quantities $\mathcal{R}_a$ or $\mathcal{S}_{ab}$, which depend only on the way the genes are replicated and inherited by daughter cells at divisions.
Using the chromosome as a reference allowed us to get rid of global fluctuations: the number of divisions considered do not affect the results, fluctuations from proteins partition at division are suppressed, all fluctuations of gene expression are cancelled, and even the division time does not appear in the final result. This argues for the assumptions that the induction time is a multiple of the division time and that the variability in division times can be neglected not to affect the results. The simulations further confirm the robustness of this strategy, in particular the values inferred with the crudest assumptions (``simple model'') are strikingly close to the true ones, both for the mean PCN per chromosome and the PCN noise (see Table \ref{tab:Sim}).

The only source of uncertainty that remains stems from the replication and partition of the plasmids and chromosome. The use of test functions suggests that it does not affect the results much. Moreover, if we suppose that both are similar, i.e. $\mathcal{R}_O\approx\mathcal{R}_G$ and $\mathcal{S}_{OO}\approx\mathcal{S}_{GG}$, we fully recover the simple model presented at the beginning and in the previous article~\cite{WongNg2010}.

The next obvious step would be to consider correlations \emph{between} cells, which could in particular inform us on plasmids partition. Here however, we lack the information on the lineage (which cells share a common induced ancestor) necessary to make a practical use of these quantities.

The use of dual reporters to dissect sources of noise was first proposed and demonstrated in a simple framework: steady state of fully induced bacteria, with both reporters in as much a similar position as possible~\cite{Swain2002,Elowitz2002}. Here we took a similar approach further, and made sense of an intuitive setup: by changing one element, namely the {\it locus} of insertion of the genes coding for fluorescent proteins, we were able to measure one particular source of noise.
The analysis proposed here could serve as a model for other derivations of this strategy.

\section*{Acknowledgements}

J.W.N. acknowledges financial support from the Minist\`ere de la Recherche and CNRS. This work was supported by the Grant No. 05-BLAN-0026-01 from the ANR.

\appendix

\section{Partition of proteins at cell divisions}
\label{App:Div}

Random partition of fluorescent proteins at cell divisions contributes only to the auto-correlation (fluctuations) of protein numbers. We suppose a binomial distribution of the number of inherited proteins. In the case of two divisions, this leads to adding the term $\frac{1}{12}\!\!\left(
\frac{7-3\mathcal{R}_a}{1+\mathcal{R}_a}\right)\!\!\langle P_a\rangle$ to $\langle P_a^2\rangle$. In turn, this translates to adding the correction 
\be
\frac{\langle P_O\rangle}{12(1+\mathcal{R}_O)}\!\!\left(\!\!\left(
\frac{1+3\mathcal{R}_O}{1+3\mathcal{R}_G}\right)\!(7-3\mathcal{R}_G)
-(7-3\mathcal{R}_O)\!\!\right)
\ee
to $\langle P_O^2\rangle$, while leaving $\langle P_G^2\rangle$ unchanged, in the expression of the PCN noise $\eta$ in the hypothesis (A)~\cite{Ghozzi2009}. This term varies from $-0.2\langle P_O\rangle$ to $0.3\langle P_O\rangle$ when we independently vary $\mathcal{R}_O$ and $\mathcal{R}_G$ in the interval $[0.15,0.45]$. Thus, with an expected number of proteins above 10 (probably hundreds to thousands), this correction is very small compared to $\langle P_O^2\rangle$ and can be safely neglected.

\section{Assumptions on gene expression and replication}
\label{App:Assump}
We make a number of assumptions in order to simplify the expression of the moments of the number of proteins at the end of induction $P_a$. They are detailed below:
\\
~
\\
\emph{(i)}  \emph{The age at the beginning of induction $t_0$, the copy numbers of plasmid or chromosome $n_a$, the rates of protein production $\alpha_a$ are independent:} the probability $p[t_0,n_a,\alpha_a]$ factorizes in $p[t_0]\,p[n_a]\,p[\alpha_a]$.
This means that there is no growth or expression burden associated with the presence of the plasmids. The notion is still debated, and whereas we could extract a small inhibition of growth for the strain bearing the mini-R1-$par^-$ plasmid, we did not see any systematic deviation: we measured comparable growth rates for all strains at a given temperature; moreover, all strains exhibited the same average green fluorescence (thus, the same average protein production)~\cite{WongNg2008}. 
\\
~
\\
\emph{(ii)} During the induction, \emph{the average protein production rate does not depend on time, on the particular copy considered or on the gene, \emph{egfp} or \emph{mOrange}}:
\be
\langle\alpha_a(i,t)\rangle\,=\,\langle\alpha\rangle,\;\forall\,a, i, t
\ee
Similarly, \emph{the correlations of two different copies do not depend on time, on the copies or on the genes}:
\be
\langle\alpha_a(i,t)\alpha_b(j,t')\rangle\,=\,\langle\alpha_O\alpha_G\rangle,\;\forall\,a,b, i\neq j, t,t'
\ee
We assume that the protein production rates immediately reach a stationary state (but this has not to hold for the proteins concentrations): Elf \emph{et al.} have shown that, at 1~mM IPTG, the fraction of LacI bound to the Lac promoter reaches its steady state value (zero) in less than 10~s~\cite{Elf2007}. Whereas the promoter we used, {\it PtacI}, is slightly different, there is no reason for the dynamics to be slower \footnote{The fact that the promoter is present in many copies even argues for these dynamics to be faster.}. 
\\
~
\\
\emph{(iii)} The dynamics of expression of {\it egfp} and {\it mOrange} are essentially fixed by the promoter; since it is the same for both genes, any copy of any of the two will follow the same statistics. Any systematic difference in the translation rate (mRNA lifetime, codon usage) is incorporated in the fluorescence per molecule factor. Since the cells are incubated overnight, with chloramphenicol blocking protein production, we expect all fluorescent proteins to have acquired their mature form. Lastly, both genes showed the same distribution of fluorescence when inserted in the chromosome~\cite{WongNg2010}.
\\
~
\\
\emph{(iv)} The gene copy numbers have reached a steady state and that there are no active loss of plasmid during the cell cycle or systematic bias in the way plasmids are inherited by daughter cells upon division: on average the plasmid and chromosome copy numbers are periodic of period $T$ and $\langle  n_a(T) \rangle=2\langle  n_a(0) \rangle$. 
\\
~
\\
\emph{(v)} On average the cross-correlations of the rates of production of proteins and of the copy numbers of two different genes do not depend on the particular copies considered nor on time for the first (we discuss different forms of the rates auto-correlation below and show that they do not affect our results much). 
\\
~
\\
\emph{(vi)} The chromosome replication and partition are perfectly controlled~\cite{Skarstad1986,Nordstroem2006}:
\be
\langle n_G(t)n_G(t')\rangle\,=\,\langle n_G(t)\rangle\langle n_G(t')\rangle.
\ee
\\
~
\\
\emph{(vii)} We approximate \emph{the plasmid copy number auto-correlation function by a constant:}
\be
\langle  n_O(t)n_O(t')\rangle-\langle  n_O(t)\rangle\langle n_O(t')\rangle=C_{n_O}\;,\forall t,t'.
\ee
This implies that $\langle  n_O(t)n_O(t')\rangle$ is periodic in each of its arguments and allows us to transform the integrals over the induction time to integrals over one cell cycle.
In the absence of a consensus model for plasmid replication or independent measurements, we cannot gauge \emph{a priori} the error we thus make. Note however that $C_{n_O}$ will not appear in the result.
\\
~
\\
\emph{(viii)} During the induction, \emph{the auto-correlation of the expression of a given copy does not depend on the gene considered, on the particular copy or on the time, but solely on the difference between two times:}
\be
\langle  \alpha_a(i,t)\alpha_a(i',t') \rangle-\langle\alpha\rangle^2=C_\alpha(|t-t'|),
\ee
where $i'$ is the \emph{ancestor} of $i$. This follows from the same arguments as given above (e.g. the dependency on $|t-t'|$ follows from the assumption that the rate of protein production reached its steady state). 

\section{Estimation of $\mathcal{R}_a$, $\mathcal{S}_{ab}$ and $\mathcal{T}_a$}
\label{App:RST}

We briefly outline here the steps allowing us to bound  $\mathcal{R}_a$, $\mathcal{S}_{ab}$ and $\mathcal{T}_a$. Full derivations can be found in~\cite{Ghozzi2009}. 

We define:
\be
\mathcal{R}_a = \frac{1}{\langle\overline{n_a}\rangle}\frac{1}{T}\int_0^T\!\!dt_0\, p(t_0)\int_0^{t_0}\!\!dt\,\langle  n_a(t) \rangle.
\ee
We linearize the age distribution: $p(t_0)=(2\ln2/T).2^{-t_0/T}\approx (2\ln2/T)(1-\ln2.t_0/T)$.
There exists $t_0^*\in [T/2,T]$ such that 
\be
\mathcal{R}_a\approx\frac{1}{\langle\overline{n_a}\rangle}\frac{2\ln2}{T^2}\left(1-\ln2\frac{t_0^*}{T}\right)\int_0^T\!\!dt_0\int_0^{t_0}\!\!dt\,\langle  n_a(t) \rangle.
\ee
We can suppose that $\langle  n_a \rangle$ is increasing on $[0,T[$. It follows that $\int_0^T\!\!dt_0\int_0^{t_0}\!\!dt\,\langle n_a(t) \rangle\leq1/2$. Recalling that $\overline{\langle n_a\rangle}=\langle\overline{n_a}\rangle$, this implies also: $\int_0^T\!\!dt_0\int_0^{t_0}\!\!dt\,\langle  n_a(t) \rangle\geq \langle  n_a(0)\rangle/(2\langle\overline{n_a}\rangle)$ and $\geq \langle\overline{n_a}\rangle/(2\langle  n_a(T) \rangle)$.
At steady state $\langle  n_a(T) \rangle=2\langle  n_a(0) \rangle$. Thus, from the preceding inequalities:
\be
\mathcal{R}_a\in[0.15,0.45].
\ee

In the same way, linearizing $p(t_0)$ and showing that $\mathcal{S}_{ab}$ can be expressed in terms of the integral of a convex function, we find $\mathcal{S}_{ab}\in[0,0.45]$.

In the case of two divisions, $\mathcal{T}_a$ is defined following:
\ba
\mathcal{T}_a&=&\frac{1}{\langle\overline{n_a}\rangle }\frac{1}{2T^2}\!\!\int_0^T\!\!dt_0 p(t_0) 
\left( 7\!\!\int_0^T\!\!dt\,t + 27\!\!\int_0^{t_0}\!\!dt\,t \right.\\
&&\left.- 3t_0\!\!\int_0^T\!\!dt + 16T\!\!\int_0^{t_0}\!\!dt- 3t_0\!\!\int_0^{t_0}\!\!dt\right)\langle n_a(t)\rangle.
\ea
We follow the same steps as before, only here we consider that each term can vary independently, thus highly overestimating the bounds for $\mathcal{T}_a$. We find $\mathcal{T}_a\in[0,9.9]$.

\section{Test functions}
\label{App:Test}

To gauge the quality of the previous estimates and fix minimal intervals, we computed $\mathcal{R}_a$, $\mathcal{S}_{ab}$ and $\mathcal{T}_a$ after postulating different shapes for the functions $\langle n_a\rangle$ and $\langle n_an_b\rangle$.

The changes of variables $t \rightarrow t/T$ and $t_0 \rightarrow t_0/T$, and the normalization 
$\langle n_a\rangle\rightarrow\langle n_a\rangle/\langle n_a(0)\rangle$ 
leave $\mathcal{R}_a$ and $\mathcal{T}_a$ unchanged. We can thus limit ourselves to increasing functions on $[0,1]$, going from $1$ to $2$.
We considered step, sigmoid, exponential, logarithmic, sinus functions and monomials of various degrees. Each type is defined by one or two parameters: each parameter was given six values in the first case and four in the second case.
 
For $\mathcal{S}_{ab}$, we considered the product of any two functions among those above. (This implies $\langle n_a(T)n_b(T)\rangle = 4\langle n_a(0)n_b(0)\rangle$, which in general is not true.)

We used the exact expression of $p(t_0)$. We found:
\be
\mathcal{R}_a^{\mathrm{test}}\!\in[0.36,0.44];
\mathcal{S}_{ab}^{\mathrm{test}}\!\in[0.20,0.28];
\mathcal{T}_a^{\mathrm{test}}\!\in[5.7,6.1].
\ee
Thus the interval found in general for $\mathcal{R}_a$ is rather good, 
whereas these results seem to confirm that the intervals found for $\mathcal{S}_{ab}$ and $\mathcal{T}_a$ were highly overestimated in the previous analysis.

\section{Simulations}
\label{App:Sim}

To test the inference method proposed in this article, we simulated roughly the experiment and compared the inferred quantities to the true ones. The results are presented in Table~\ref{tab:Sim}. We introduce here simple models of gene expression and replication, but recall that no such models are assumed in the inference method. The transcription of each reporter gene, the translation of the corresponding RNAs and their degradation, and the replication of the plasmids are each implemented as single stochastic reactions~\cite{Anderson2007}:
\ba
\mathrm{Gene}_a + \mathrm{Inducer} & \longrightarrow & \mathrm{Gene}_a + \mathrm{Inducer} +  \mathrm{mRNA}_a\\
\mathrm{mRNA}_a & \longrightarrow & \mathrm{mRNA}_a + \mathrm{Protein}_a\\
\mathrm{mRNA}_a & \longrightarrow & \emptyset\\
\mathrm{Gene}_a & \longrightarrow & 2\,\mathrm{Gene}_a
\ea
where $a$ again refers to either the plasmid or the chromosome. 
We follow here a simplified reactions scheme and ignore the polymerase binding to the promoter, the formation of an open complex and initiation of transcription, the binding of ribosomes on mRNAs, 
as well as the formation of complexes with RNases before mRNAs' decay. We used kinetic parameter values such as to find the same  protein number distributions (expressed from the chromosome) as with the full reaction scheme simulated in~\cite{Swain2002}, and Swain et al. found a good agreement between the two schemes: the parameters of the simple scheme effectively catch these underlying processes.

Noise sources related to the elongation of mRNAs and proteins (the fact that they are not produced in single steps) are also ignored. They would include the traffic of polymerases or ribosomes, pauses in transcription or translation, etc. We guess that they could similarly be accounted for in the rate constants and thus would not qualitatively change the results.

The rates of transcription and translation are the same for both reporters.
We impose the number of chromosomes to go from 1 to 2 at 40\% of the cell cycle. The number of inducers is also imposed: it is 0 until some time $t_\mathrm{lag}$, and 1 until the end of the simulation. 

The duration of the cell cycle $T$ is either fixed (cases 1 and 2) or drawn from a gamma distribution with parameters chosen so that the average division time is 1800 or 1200 s, and typical variations of 10\% (cases 3 and 4)~\cite{Lebowitz1974}. The age of the cell at the beginning of the simulation is either 0 (case 1) or drawn from the exponential distribution indicated in the main text. The time $t_\mathrm{lag}$ has been arbitrarily fixed at 10 times the mean cell cycle duration. The induction, i.e. the remainder of the simulation, has always been taken to last 3600 s.

We follow, in each simulation, one lineage. Upon cell division, the RNAs, proteins and plasmids are randomly picked to stay (their numbers drawn from a binomial distribution). 100,000 simulations were performed in each of the four cases.

Importantly, there is no control of the plasmid copy number: we simply fix the rate of replication to the growth rate $\ln2/T$, so that the plasmids are on average replicated once per cell cycle. A steady state of plasmid copy number is thus not reached, contrary to the more realistic assumption made earlier; the fact that we can still recover the mean PCN and the PCN noise shows that even this assumption is not critical. 

Each cell has 10 plasmids at the beginning of the simulation.
The cell cycle average $\overline{\bullet}$ is taken during the first full cycle of induction.

All codes are available upon request.


\bibliography{PlasmidesTheo.bib}

\end{document}